\documentstyle[pra,aps]{revtex}
\begin{document}
\draft
\title{Small Denominators, Frequency Operators, 
       and Lie Transforms for Nearly-Integrable Quantum Spin 
       Systems\cite{diplom}}
\author{Thomas Gramespacher and Stefan Weigert}
\address{Institut f\"ur Physik der Universit\"at Basel,
	Klingelbergstrasse 82, CH-4056 Basel}
\date{\today}
\maketitle
\begin{abstract}
Based on the previously proposed notions of action operators and of 
quantum integrability, {\em frequency operators\/} are introduced in a 
fully quantum-mechanical setting. They are conceptually useful because a new 
formulation can be given to unitary perturbation theory.  When worked out 
for quantum spin systems, 
this variant is found to be formally equivalent to canonical perturbation 
theory applied to nearly-integrable systems consisting of classical spins. 
In particular, it becomes possible to locate the quantum-mechanical 
operator-valued equivalent of the {\em frequency denominators\/} 
which may cause divergence of the classical perturbation series. The results 
established here link the concept of quantum-mechanical integrability 
to a technical question, namely the behaviour of specific perturbation series.
\end{abstract}
\pacs{03.65.Bz,03.65.Fd,03.65.Ca}
\section{Introduction}
In classical mechanics, the distinction between integrable and nonintegrable
Hamiltonian systems is clear-cut. On the one hand, for a system to be 
integrable, the existence of $N$ smooth, involutive, and functionally 
independent constants of the motion is required. Then, its $2N$-dimensional 
phase space is guaranteed to be foliated into $N$-dimensional tori, and  
the resulting almost-periodic motion of the system is very simple. On the 
other hand, a nonintegrable system may explore a $(2N-1)$-dimensional 
region of phase space in the course of time if (except for the energy) there 
are no invariants which restrict the motion to lower-dimensional manifolds. 
Accordingly, the time evolution of a nonintegrable system is extremely 
complicated \cite{lichtenberg+83}. 

Much is known about the properties of 
quantum systems obtained by quantizing classically integrable and 
nonintegrable systems, respectively \cite{giannoni+91}. Nevertheless,
no rigorously established and generally accepted features have emerged
which would allow one to unambiguously distinguish integrable and 
nonintegrable quantum systems in analogy to classical mechanics. 
While the concept of integrability is important in classical mechanics (for 
example, the KAM theorem deals with the influence of a perturbation on 
{\em integrable\/} systems), it appears, for the time being, to be much less 
fertile in quantum mechanics. 

A new concept of {\em quantum integrability\/} has been proposed in 
\cite{weigert+95} which is based on quantum-mechanical {\em action 
operators\/} without any reference to classical mechanics. 
Incorporating criticism of earlier concepts of quantum integrability 
\cite{weigert92}, this notion relies only on the {\em algebraic\/} structure 
of a quantum system  provided by the commutation relations of the basic 
operators. It is the objective of the present paper to explore the features 
of nearly quantum integrable systems: a perturbation is added to a system 
which qualifies as quantum integrable in the above sense, and a method is 
given to successively remove the disturbance. By 
using the quantum actions in a framework of Lie transforms it becomes 
possible to lay open strong formal analogies to the corresponding 
classical perturbational approach. In particular, with {\em frequency 
operators\/} being defined for integrable quantum systems, 
it will become obvious that quantum-mechanical ``counterparts" 
of the classical {\em frequency denominators\/} show up in this
perturbational treatment. The results of the present work are 
{\em formal\/}, i.e., questions of convergence have not yet been dealt with. 
Fortunately, the structural similarity of the results to classical mechanics
allows one already to imagine possible scenarios. 

Two general remarks are in order before turning to the 
detailed presentation. The investigations are carried out 
for {\em spin\/} systems but the generalization of the results to 
particle systems is not expected to provide substantial 
difficulties. Then, it should be pointed out that no part of this work 
relies on semiclassical approximations.
 
In brief, the paper is 
organized as follows. In Section 2, the notion of quantum 
integrability used here is recapitulated. Frequency operators are 
introduced in Section 3, and they are recognized to 
naturally generalize the classical frequencies known from integrable 
systems. The main part of the paper, Section 4, consists in formulating 
the theory of Lie transforms for nearly quantum integrable systems. 
Applications to systems with one and two degrees of freedom follow in 
Section 5. Finally, the results are summarized in Section 6, 
and connections to other work are discussed.
\section{The algebraic view of quantum integrability}
It will be useful to summarize the algebraic view of quantum integrability 
in order to motivate the investigations of the subsequent chapters. Only 
the essential points will be mentioned; details have been elaborated on in 
Ref. \cite{weigert+95}. 

Let us outline the concept of integrability with a view on the algebraic 
structure of classical mechanics. For a system of $N$ degrees of 
freedom there are canonical variables $q = (q_{1}, \ldots, q_{N})$ and 
$p = (p_1, \ldots, p_{N})$. They provide a basis for the algebra of 
(smooth) phase-space functions $f(p, q)$ since each such function can be 
approximated arbitrarily well 
by summing products of $p_{n}, q_{n}$ with appropriate coefficients. 
They can be multiplied by real coefficients, and one can 
add them; any two phase-space functions define a third one  
via the Poisson bracket turning the set of all phase-space functions into 
a Lie algebra. This framework is the same for all 
classical mechanical systems. A particular physical system is described by 
selecting 
one specific phase-space function and by calling it the Hamiltonian. Its 
r\^ole is to generate the time evolution of the system by defining a flow in 
phase space. Integrability comes in as follows. Typically, the Hamiltonian 
function will depend on all components of momentum $p$ and 
coordinate $q$. However, it is possible to perform a canonical change of 
basis, $(p, q) \to (p', q')$, as a result of which the Hamiltonian will 
turn into a different function if expressed in terms of the primed 
variables. It may happen that such a change of basis leads to a particular 
form of the new Hamiltonian function, namely that it depends on only half 
of the number of variables, $N$. One has found a particularly 
convenient basis of the algebra because it becomes obvious that the flow 
in phase space decomposes into $N$ decoupled flows. A system with such a 
{\em trivial\/} flow is called {\em integrable\/}. As is well-known, this 
situation is very rare: typically no decoupling basis exists, and the 
phase-space flow is extremely intricate. 

It is straightforward
to rephrase this line of thought for quantum-mechanical systems. Now the 
basis of the algebra of operators is given by $\widehat p= (\widehat p_{1}, 
\ldots, \widehat p_{N})$ and $\widehat q = (\widehat q_{1}, \ldots, 
\widehat q_{N})$, and 
each operator in the algebra can be represented by an appropriate sum of 
products of $\widehat p_{n}, 
\widehat q_{n}$. The commutator of two operators defines a third one, and 
Jacobi's identity holds in general: one is dealing with a Lie algebra of 
operators. If one particular operator of 
the algebra is chosen as Hamiltonian, a quantum system and its time 
evolution are determined. A generic Hamiltonian operator will depend  on 
all components of $\widehat p$ and $\widehat q$. In special cases, however, a 
{\em unitary\/} transformation exists which allows one to express the 
Hamiltonian as a function of only half of the number of basic operators, $N$. 
By analogy, the system will then be called {\em quantum integrable\/}. Part of 
the purpose of the present work is to show that, in an appropriate sense, 
the corresponding quantum-mechanical flow is {\em trivial\/}, and that this 
situation is not generic. 

It should be noted that in the context of quantum mechanics the term 
``algebraic''  will be understood as ``independent of the representation.'' 
The idea is to only 
use relations which have a counterpart in the classical algebra.  
For example, within each finite-dimensional matrix representation of the 
spin algebra there exist relations between different powers of 
spin matrices which are specific for the representation at hand. Such 
relations do not exist in the classical algebra and, thus, they should not 
be used. An instructive discussion of the common algebraic 
structure underlying classical and quantum mechanics has been given by Falk 
\cite{falk51}.

For convenience, the definition of quantum integrability  for spin 
systems proposed in \cite{weigert+95} will be reproduced.
The description of a quantum spin system with $N$ degrees of freedom 
is based on $N$ copies of the one-spin commutation relations,
\begin{equation}
[\widehat S_{j\alpha}, \widehat S_{j^\prime\beta}]
	= i \hbar \delta_{jj^\prime} \sum_{\gamma = x y z}
		  \varepsilon_{\alpha \beta \gamma} \widehat S_{j\gamma}
	 \, ,
	 \qquad
	 j,j^\prime = 1,2, \ldots N \, .
\label{Ncommrel}
\end{equation}
Here is the criterion for integrability of quantum spin systems: 
\begin{itemize}
\item[QJ:] 
A given $N$-spin Hamiltonian
$\widehat{ H} 
    = H ( \widehat{\bf S}_1, \ldots ,\widehat{\bf S}_N )$
is {\em quantum integrable\/} if there exists a unitary transformation
${\cal U} (\widehat{\bf S}_1, \ldots ,\widehat{\bf S}_N)$ 
which converts the spin operators
$\mbox{$\widehat{\bf S}$}_j$, $j=1,\ldots , N $, into new spin operators,
\begin{equation}
\widehat{\cal S}_j
       = \widehat{\cal U} \, \widehat{\bf S}_j \, \widehat{\cal U}^\dagger
       = {\cal S}_j (\widehat{\bf S}_1, \ldots \widehat{\bf S}_N) \, ,
	\qquad j = 1, \ldots , N \, ,
\label{Njunitary}
\end{equation}
such that the Hamiltonian turns into a function of $N$ quantum actions 
$\hat J_j = \widehat{\cal S}_{jz}$: 
\begin{equation}
H ( \widehat{\bf S}_1, \ldots , \widehat{\bf S}_N)
	   = {\cal H} (\hat{J}_{1}, \ldots, \hat{J}_{N}) \, .
\label{Htrf}
\end{equation}
The spectra of the operators $\hat{J}_{j}, j=1, \ldots ,N$,
in each $(2s+1)$-dimensional representation $\Gamma_s$ consist of uniformly 
spaced levels:
\begin{equation}
\hat J^{s}_{j}
     = \sum^{+s}_{m_j = -s}
	  |  \ldots , m_j ,\ldots ; s  \rangle  \, 
	  m_j \hbar \, 
	  \langle  \ldots, m_j ,\ldots ; s | \, , 
	  \quad  j=1, \ldots ,N \, .
\label{qmactionN}
\end{equation}
\end{itemize}
If the transformation (\ref{Htrf}) has been achieved, the eigenvalues of the 
Hamiltonian can be read off immediately, and its eigenfunctions are known 
as well. In spite of reference to the representation $\Gamma_s$, 
Eq. (\ref{qmactionN}) is an algebraic statement since it is required to hold 
in {\em all \/} representations, $s=1/2, 1,3/2, \ldots$ As will become 
clear in Section 4, condition (\ref{Htrf}) defines a program which can be 
carried out in strong analogy to classical mechanics.    
\section{Frequency operators}
Frequency operators, to be defined for quantum integrable systems in a 
straightforward way, are introduced in this section. The 
main objective of Ref. \cite{weigert+95} has been to provide numerical 
and analytical support for the existence of  quantum-mechanical action 
operators, and to use them as building blocks for quantum-mechanical 
integrability. These arguments will not be repeated here---instead   
an independent approach will be presented which illustrates the idea of 
action operators from a new perspective. To arrive at the definition of 
frequency operators will then be a small step.  

As a starter, a classical spin system with one degree of freedom is 
considered. Its Hamiltonian is assumed to depend on a single spin 
component only, $S_z$, say:
\begin{equation}
H=H(S_z)\, . 
\label{clham}
\end{equation}
The complex quantities
\begin{equation}
S_\pm = S_x \pm iS_y
\label{clladder}
\end{equation}
are classical counterparts of ladder operators for a quantum spin. In 
these variables the Poisson brackets for a single spin read 
\begin{equation}
\{ S_z , S_\pm \} = \pm i S_\pm \, , \qquad\{S_+,S_-\} = 2 i S_z \, ,
\label{clpol}
\end{equation}
as obtained from the replacement $[ \, , \, ] \to -i\hbar \{ \,  , \, \}$   
in (\ref{Ncommrel}). The equations of motion are given by 
\begin{equation}
\frac{d {\bf S}}{dt} = \{ H, {\bf S} \} \, ,
\label{classeom}
\end{equation}
and for the Hamiltonian (\ref{clham}) they read explicitly
\begin{eqnarray}
\frac{dS_z}{dt} & = & \{ H, S_z \} = 0 \, ,\\
\frac{dS_+}{dt} & = & \{ H, S_+ \} = i S_+ \, \omega(S_z) \, , 
\label{cleqmo}
\end{eqnarray}
the equation of motion for $S_-$ being complex conjugate to that 
of $S_+$. When expressing the spin as 
${\bf S} =  ( \sqrt{S^2-p_\varphi^2} \cos \varphi , 
	      \sqrt{S^2-p_\varphi^2} \sin \varphi , 
	      p_\varphi ) $ 
in terms of canonical coordinates with  $\{ p_\varphi, \varphi \} = 1$, it 
becomes obvious that the third component, $S_z$, plays the 
r\^ole of an action: it is both a canonical variable and a constant of the  
motion. The phase-space functions $S_\pm$ are proportional to the 
exponentiated angle variable. In view of later developments it is useful to 
stick with the periodic functions of time, $S_\pm(t)$, instead of using the 
canonical variable $\varphi$. The frequency
\begin{equation}
\omega(S_z)=\frac{dH(S_z)}{dS_z} 
\label{clfreq}
\end{equation}
is completely determined by the Hamiltonian and thus depends on the 
action $S_z$ only. The time evolution of integer powers $k = 2,3, \ldots$ 
of $S_+(t)$ is obtained from (\ref{cleqmo}) as 
\begin{equation}
\frac{d S_+^k }{dt} = \{ H, S^k_+ \} = i S_+^k \, \omega_k (S_z) \, ,
\label{clpowers}
\end{equation}
and similarly for $S_-$. The notation 
\begin{equation}
\omega_k(S_z) \equiv k \omega(S_z) \, , 
\label{clfrpower}
\end{equation}
used here will simplify comparison with 
quantum-mechanical expressions.

Consider now a quantum-mechanical spin system with one degree of 
freedom, i.e., one copy of the commutation relations (\ref{Ncommrel}) 
only occurs. The Hamiltonian of the system is assumed to depend 
on the third component of the spin operator $\widehat{\bf S}$ only,
\begin{equation}
\widehat{H} = H (\widehat{S}_z)\, . 
\label{qmham}
\end{equation}
Again, ladder operators are useful: 
$\widehat{ S}_\pm = \widehat{ S}_x \pm i\widehat{ S}_y,$ leading to 
commutators similar to (\ref{clpol}):
\begin{equation}
[ \widehat{ S}_z , \widehat{ S}_\pm ] 
	       = \pm \hbar \widehat{ S}_\pm \, , 
\qquad 
[ \widehat{ S}_+, \widehat{ S}_- ] 
	       = 2 \hbar \widehat{ S}_z \, . 
\label{qmpol}
\end{equation}
Heisenberg's equations of motion for the quantum spin are
\begin{equation}
\frac{d \widehat{ \bf S}}{dt} 
  = \frac{i}{\hbar} [ \widehat{ H}, \widehat{ \bf S} ] \, ,
\label{qmeom}
\end{equation}
and for the Hamiltonian (\ref{qmham}) they read
\begin{eqnarray}
\frac{d\widehat{S}_z}{dt} & = &  
		 \frac{i}{\hbar} [ \widehat H, \widehat{S}_z ] 
		 = 0 \, , \\
\frac{d\widehat{S}_+}{dt} & = & 
		 \frac{i}{\hbar} [ \widehat H, \widehat{S}_+ ] 
		 = i \widehat{S}_+ \, \omega (\widehat S_z) \, ,
\label{qmeqmo}
\end{eqnarray}
and the equation for $\widehat{S}_- = (\widehat{S}_+)^\dagger$ is the 
adjoint of the last equation,
\begin{equation}
\frac{d\widehat{S}_-}{dt}
	 = \frac{i}{\hbar} [ \widehat H, \widehat{S}_- ] 
	 = - i \omega(\widehat S_z) \,  \widehat{S}_- \, .
\end{equation}
The selfadjoint operator 
$\widehat{\omega} \equiv \omega(\widehat{S}_z)$ 
depends on the $z$ component of the spin operator $\widehat{\bf S}$ only; 
explicitly, one has
\begin{equation}
\omega(\widehat{S}_z) = \frac{1}{\hbar} \left( H(\widehat{S}_z+\hbar)
		   - H(\widehat{S}_z)\right)\, ,
\label{qmfreq}
\end{equation}
providing thus the definition of a quantum-mechanical {\em frequency 
operator\/}. Its form is suggested by requiring that the operator 
$\widehat{S}_+$ 
stand on the left of everything else in Eq. (\ref{qmeqmo}). 
The derivation of (\ref{qmfreq}) hinges on the fact that 
the Hamilton operator depends on the operator $\widehat{S}_z$ only because 
in (\ref{qmeom}) one can then use the relation 
\begin{equation}
[ f ( \widehat{ S}_z ) , \widehat{ S}_+^k ] 
   = \widehat{ S}_+^k \left(f ( \widehat{S}_z + k \hbar)  
	     - f ( \widehat{S}_z ) \right) 
 = \widehat{ S}_+^k \Big( \exp ( k \hbar \partial_z ) - 1 \Big) 
	     f ( \widehat{S}_z )  \, ,
\label{niceformel}             
\end{equation}
where $\partial_z$ denotes the (formal) derivative with respect to  
the operator $\widehat{ S}_z$. An algebraic proof of (\ref{niceformel}) is 
given in Appendix A for functions $f(x)$ which have a power-series expansion 
in $x$, and another one for arbitrary functions $f(x)$ which holds in  
each representation. A different but equivalent definition of the frequency 
operator can be given if one moves $\widehat{S}_+$ to the right in Eq. 
(\ref{qmeqmo}). Either convention is unambiguous; in the 
following, the version occurring in (\ref{qmeqmo}) will be used. 

Comparing the classical and the quantum-mechanical equations of 
motion, (\ref{cleqmo}) and (\ref{qmeqmo}), it becomes plausible to 
call the operator $\widehat{S}_z$ an {\em action operator\/}: it is both a  
constant of the motion and (the closest analog of) a canonical momentum 
in the spin algebra. It is reasonable to consider the Hamiltonian 
(\ref{qmham}) as {\em quantum integrable\/} since it depends on nothing 
but an action  operator. This also fits well with conceiving 
$\widehat{\omega}$ given in (\ref{qmfreq}) as frequency operator because  
(\ref{qmfreq}) is the discretized version of (\ref{clfreq}).

On this basis, the notion of quantum-mechanical integrability as 
defined in QJ makes sense: a system with Hamilton 
operator $\widehat{\cal H}$ will be called quantum integrable if a unitary 
transformation $\widehat{\cal U}$ can be found such that the 
Hamiltonian turns into a function of commuting $z$ components of the 
different spins only while the commutation relations (\ref{Ncommrel}) remain 
invariant. As a consequence, the equations of motion in the Heisenberg 
picture  strongly resemble those of classically integrable 
systems. In addition, frequency operators are defined in analogy to 
the classical frequencies both of which are determined completely by the 
Hamiltonian. It is not for the first time that operator-valued analogs of 
the classical frequencies do occur: as early as 1925, Dirac introduced a 
similar concept for particle systems \cite{dirac25}.

It is important to look at the time evolution of powers of $\widehat{S}_+$ 
(and $\widehat{S}_-$). One finds
\begin{equation}
\frac{d \widehat{S}_+^k}{dt} 
	 = \frac{i}{\hbar} [ \widehat H, \widehat  S^k_+ ]
	 = i \widehat{S}_+^k \, \omega_k (\widehat S_z) \, ,
\label{powereq}
\end{equation}
where now
\begin{equation}
\omega_k(\widehat{S}_z) 
	= \frac{1}{\hbar} \left( H (\widehat{S}_z + k \hbar)
	     - H ( \widehat{S}_z ) \right)  \, .
\label{qmfreqpower}
\end{equation}
Since generally
\begin{equation}
\widehat{\omega}_k \neq k \widehat{\omega} \, , 
\label{disparity}
\end{equation}
a difference to the classical relation (\ref{clfrpower}) arises which will 
have interesting consequences for the perturbative approach to be studied 
later on. When formally expanding both sides of (\ref{disparity}) in powers 
of $\hbar$, however, the leading order reflects the classical relation. 

The generalization to systems consisting of $N$ spins, 
$\widehat{ \bf S}_{j}, j=1,\ldots ,N$, is staightforward. The 
components of each individual spin fulfill commutation relations
(\ref{Ncommrel}), and operators associated with different spins $\widehat 
S_{j\alpha}, \widehat S_{j^\prime \beta}$ commute. The Hamiltonian now 
will depend on all spins, $\widehat  H 
= H (\widehat{ \bf S}_{1},\ldots, \widehat{ \bf S}_{N})$ and the equations 
of motion read
\begin{equation}
\frac{d\widehat S_{j \alpha} }{dt} 
	= \frac{i}{\hbar} [\widehat H, \widehat  S_{j \alpha}] \, , 
	\qquad \alpha = x, y, z \, .
\label{Neqsmo}
\end{equation}
For a quantum integrable $N$-spin Hamiltonian there are $N$ frequency 
operators generalizing the expressions (\ref{qmfreq}) and (\ref{qmfreqpower}):
\begin{equation}
\omega_{k}^{(j)} ( \widehat{ \bf S}_{z}) 
     = \frac{1}{\hbar} \left( H (\widehat S_{1z}, \ldots, 
	     \widehat S_{jz} + k \hbar, \ldots ,\widehat  S_{Nz}) 
    - H ( \widehat{ \bf S}_{z}) \right) \, ,
\label{Nfreq}
\end{equation}
and the time evolution of the ladder operators is simply given by 
\begin{equation}
\frac{d \widehat  S_{j+}^{k}}{dt}  
      = \frac{i}{\hbar} [ \widehat H, \widehat  S^k_{j+} ]    
      = i \widehat  S^{k}_{j+} \widehat  \omega^{(j)}_{k} \, , 
      \quad k = 1, 2 , \ldots \, ,
\label{heisenbergopeq}
\end{equation}
and their adjoints.
\section{Perturbation Theory}
Having established a concept of integrability for quantum systems,
the program to be carried out is as follows. In classical mechanics, a 
small perturbation is added to an integrable system and it is 
investigated whether, by successive canonical transformations, the 
perturbation can be removed completely; this would render the system 
integrable. In consequence of the KAM theorem, however, integrability is 
destroyed by a generic perturbation: any perturbation series is doomed to 
diverge in finite fractions of phase space.  A quantum-mechanical 
implementation of this approach aims at transforming a perturbed quantum 
integrable system into an integrable one by applying a sequence of 
unitary transformations. Lie transforms are a convenient tool here since 
they are adapted to the algebraic structure of the theory. Moreover, 
they facilitate comparison with canonical perturbation theory, the 
relevant formulae of which are gathered in Appendix B. 

A quantum system  consisting of $N$ spins with constant length 
$\widehat S_j^2$ is assumed to be described by a Hamiltonian operator 
of the form 
\begin{equation}
\widehat{H}_\varepsilon=H(\widehat{\bf S},\varepsilon) 
   = H_0(\widehat{\bf S}_z) + \widetilde H(\widehat{\bf S},\varepsilon)\, ,
\label{generalform}
\end{equation}
where $\widehat{H}_0$ is a quantum integrable system, depending only on the 
$z$ components of the spins (if not indicated otherwise, $\widehat{\bf S}$ 
denotes the collection $(\widehat{\bf S}_1 , \ldots , \widehat{\bf S}_N)$ of 
operators).

The perturbation 
$\widetilde{H}(\widehat{ \bf S}, \varepsilon)$ may depend on all components of 
the $N$ spins and it contains a real expansion parameter $\varepsilon$ 
such that the perturbation vanishes for $\varepsilon=0$: 
$\widetilde{H}(\widehat{ \bf S}, 0) = 0$. Furthermore, it is assumed that 
the unperturbed Hamiltonian $\widehat{H}_0$ is not degenerate. In the 
following, a possible dependence of operators on the invariants 
$\widehat S_j^2 $ will not be made explicit. 
\subsection{Unitary transformations}
To begin with, it is useful to study the effect of unitarily transforming 
the spin operators by an operator 
$\widehat{U}_\varepsilon = U (\widehat{\bf S},\varepsilon)$
with $U (\widehat{\bf S}, \varepsilon=0) = 1$.   
New spin operators $\widehat{\bf S}'$ (throughout this paper the prime 
does not denote a derivative) are introduced by
\begin{equation}
\widehat{\bf S}'_\varepsilon 
  = U_\varepsilon (\widehat{\bf S}) \, 
      \widehat{\bf S}               \,
    U_\varepsilon^\dagger (\widehat{\bf S}) \, ,
\label{spintrf}
\end{equation}
and the induced transformation of the Hamiltonian reads
\begin{equation}
H(\widehat{\bf S},\varepsilon) 
       = H'(\widehat{\bf S}'_\varepsilon, \varepsilon)\, ,
\label{hamtrf}
\end{equation}
that is, in terms of the primed operators, $\widehat{\bf S}'_\varepsilon$, 
the Hamiltonian is given by a new function $H'$. By using the inverse of 
(\ref{spintrf}) in the form
\begin{equation}
\widehat{\bf S} 
  = U_\varepsilon^\dagger (\widehat{\bf S}) \, 
	     \widehat{\bf S}_\varepsilon^\prime  \,
    U_\varepsilon (\widehat{\bf S})  
  = U_\varepsilon^{\prime \dagger}            
			      (\widehat{\bf S}_\varepsilon^\prime) \,
	     \widehat{\bf S}_\varepsilon^\prime    \,
    U_\varepsilon^\prime (\widehat{\bf S}_\varepsilon^\prime)  \, ,
\label{inversespintrf}
\end{equation}
relation (\ref{hamtrf}) can be written as
\begin{equation}
U_\epsilon^{\prime \dagger} (\widehat{\bf S}_\varepsilon^\prime ) \, 
	  H (\widehat{\bf S}_\varepsilon^\prime, \varepsilon)    \,
	      U_\epsilon^\prime (\widehat{\bf S}_\varepsilon^\prime )
=  H^\prime (\widehat{\bf S}'_\varepsilon, \varepsilon) \, .
\label{generalhamtrf}
\end{equation}
This equation holds for any unitary transformation $\widehat{U}_\varepsilon$. 
In view of the criterion for quantum integrability, QJ, the new Hamiltonian 
$\widehat{ \cal H} = \widehat{H}'$ is required to depend on the $z$ components 
of the primed spin operators $\widehat{\cal S} = \widehat{\bf S}^\prime $ 
only, so that (\ref{hamtrf}) assumes the special form
\begin{equation}
H(\widehat{\bf S},\varepsilon) 
	   = {\cal H}(\widehat{\cal S}_z,\varepsilon) \, . 
\label{qinthamtrf}
\end{equation}
Defining $ {\cal U}_\varepsilon (\widehat{\cal S}) 
	  = U_\varepsilon^\prime (\widehat{\cal S})$,
Eq. (\ref{generalhamtrf}) turns into
\begin{equation}
{\cal U}_\epsilon^\dagger (\widehat{\cal S} )  \,
	  H (\widehat{\cal S}, \varepsilon)        \,
	      {\cal U}_\epsilon (\widehat{\cal S} )
=  {\cal H}(\widehat{\cal S}_z , \varepsilon) \, .
\label{pluginhamtrf}
\end{equation}
In order to distinguish quantities related to a general transformation 
as in (\ref{hamtrf}) from those associated with (\ref{qinthamtrf}), curly 
symbols such as $\widehat{\cal U}$ and $\widehat{\cal H}$  
will be used. It will be convenient to rename the curly operators 
$\widehat{\cal S}$ by $\widehat{\bf S}$ in Eq. (\ref{pluginhamtrf}). 
\subsection{Perturbation expansion}
The construction of the unitary transformation $\widehat{\cal U}_\varepsilon$ 
achieving (\ref{pluginhamtrf}) is at stake now.  Under three assumptions 
it can be reduced formally to the successive solution of an infinite set of 
hierarchical equations determined by the unperturbed Hamiltonian 
$H_0(\widehat{\bf S}_z)$ and the perturbation 
$\widetilde{H}(\widehat{ \bf S },\varepsilon)$. First, it is assumed that the 
perturbation in (\ref{generalform}) can be expanded in a power series of the 
parameter $\varepsilon$:
\begin{equation}
\widetilde{H} (\widehat{\bf S},\varepsilon) 
	 = \sum_{p=1}^\infty \varepsilon^p H_p(\widehat{\bf S})\, .
\label{pertpower}
\end{equation}
When writing the operator $\widehat{\cal U}_\varepsilon$ as a Lie transform,
\begin{equation}
{\cal U}(\widehat{\bf S},\varepsilon) 
	   = \exp(-iu(\widehat{\bf S},\varepsilon)/\hbar)\, ,
\label{expu}
\end{equation}
a similar expansion is required to exist for the hermitean 
operator $\widehat{u}_\varepsilon$:
\begin{equation}
u(\widehat{\bf S},\varepsilon) 
	= \sum_{p=1}^\infty \varepsilon^p u_p(\widehat{\bf S})\, .
\label{upower}
\end{equation}
The operator $\widehat{u}_\varepsilon$ is called the generator (or generating 
operator) of the unitary transformation $\widehat{\cal U}_\varepsilon$. 
Finally, it must be possible to expand the new Hamiltonian operator, 
$\widehat{\cal H}$, in powers of $\varepsilon$:
\begin{equation}
{\cal H}(\widehat{S}_z ,\varepsilon) 
	  =  \sum_{p=0}^\infty \varepsilon^p {\cal H}_p (\widehat{S}_z) 
		       \, .
\label{newhampowers}
\end{equation}
Plugging these expansions into 
requirement (\ref{pluginhamtrf}) and collecting terms multiplied by equal 
powers of the parameter $\varepsilon$, one obtains the following set of 
equations:
\begin{eqnarray}
0 
& = & \widehat{H}_0-\widehat{\cal H}_0 \, , \nonumber \\
\frac{i}{\hbar}[\widehat{H}_0,\widehat{u}_1] 
	       & = & \widehat{H}_1-\widehat{\cal H}_1 \, , \nonumber \\
\frac{i}{\hbar}[\widehat{H}_0,\widehat{u}_2] 
	       & = & \widehat{H}_2 
		     - \frac{i}{2\hbar}[\widehat{u}_1,\widehat{\cal H}_1 
		     + \widehat{H}_1] - \widehat{\cal H}_2 \, , \nonumber \\
	       & \vdots & \nonumber \\
\frac{i}{\hbar}[\widehat{H}_0,\widehat{u}_p] 
	       & = & R_p(\widehat{u}_1,\ldots,\widehat{u}_{p-1},\widehat{H}_1,
			    \ldots, \widehat{H}_p,\widehat{\cal H}_1,\ldots,
			    \widehat{\cal H}_{p-1}) - \widehat{\cal H}_p  
			    \, ,  \nonumber \\
	       & \vdots &    
\label{hierarchy1}
\end{eqnarray}
For convenience, each equation has been solved 
for the commutators of the unperturbed Hamiltonian, $\widehat{H}_0$, with the 
generator of the highest index, $\widehat{u}_p$. Eqs. (\ref{hierarchy1}) 
constitute a nested hierarchy: at the $p$th level the quantities 
$\widehat{\cal H}_p$ and $\widehat{u}_p$ can be determined if all quantities 
with smaller indices have been found; the operator $\widehat{ R}_p$ contains 
only quantities know from the beginning or determined when solving the 
equations of the lower levels. Such a structure is typical for 
perturbation theory formulated in terms of Lie transforms. It is possible to  
discuss the method to solve the equations (\ref{hierarchy1}) in full 
generality \cite{cary81,shavitt+80,sibert88}. For simplicity, the procedure 
is carried out here for  a system with a single degree of freedom first; 
generalization to two (or more) spins is then straightforward.
\subsection{One degree of freedom}
For a single-spin system with Hamiltonian
\begin{equation}
H (\widehat{\bf S},\varepsilon) 
   = H_0(\widehat{S}_z) + \widetilde H(\widehat{\bf S},\varepsilon)\, ,
\label{generalform1}
\end{equation}
the unknowns $\widehat u_p$ and $\widehat{ \cal H}_p$ are determined from 
the nested hierarchy (\ref{hierarchy1}) as follows.

Solving the first equation of the hierarchy is particularly simple because it 
states that the old and the new Hamiltonian, $\widehat{H}_0$ and 
$\widehat{\cal H}_0$, respectively, coincide. By assumption, $\widehat{H}_0$ 
depends on the $z$ component of $\widehat{\bf S}$ only, and, thus, 
$\widehat{\cal H}_0$ has this property, too, while $\widehat{\cal S}$ and 
$\widehat{\bf S}$ are identical to order $\varepsilon^0$. This is  
consistent as is easily seen from the limit of vanishing perturbation,  
$\varepsilon = 0$.

The remaining equations are all of the same type:
\begin{equation}
\frac{i}{\hbar}[H_0(\widehat{S}_z), u_p(\widehat{S}_\pm,\widehat{S}_z)] 
       = R_p(\widehat{S}_\pm,\widehat{S}_z) - {\cal H}_p ( \widehat{S}_z) \, .
\label{integrate1}
\end{equation}
The operator $\widehat{R}_p$ can be written as a power series in 
the operators $\widehat{S}_\pm$ (cf. Eq. (\ref{orderF1}) in Appendix C):
\begin{equation}
R_p(\widehat{S}_\pm,\widehat{S}_z)
	  = R_{p,0}(\widehat{S}_z)
	     + \sum_{k=1}^\infty 
	     \left(\widehat{S}_+^k R_{p,k}^+(\widehat{S}_z) 
	     + R_{p,k}^-(\widehat{S}_z)\widehat{S}_-^k
	     \right) \, , 
\label{expandr}
\end{equation}
with uniquely defined operators 
$\widehat{R}_{p,k}^+ = (\widehat{R}_{p,k}^-)^\dagger$ 
and $\widehat{R}_{p,0}$ which depend on $\widehat{S}_z$ only. 

On the other hand, expanding $u_p(\widehat{S}_\pm,\widehat{S}_z)$ as in 
(\ref{expandr}), 
\begin{equation}
u_p(\widehat{S}_\pm , \widehat{S}_z)
	  = u_{p,0} (\widehat{S}_z)
	     + \sum_{k=1}^\infty 
	     \left(\widehat{S}_+^k u_{p,k}^+(\widehat{S}_z) 
	     + u_{p,k}^-(\widehat{S}_z)\widehat{S}_-^k
	     \right) \, ,
\label{expandu}
\end{equation}
one obtains for the commutator in (\ref{integrate1})
\begin{equation}
\frac{i}{\hbar}[\widehat{H}_0,\widehat{u}_p]
	    = \sum_{k=1}^\infty
	      \left(\widehat{S}_+^k \widehat{u}_{p,k}^+ i\widehat{\omega}_k
		   + (-i \widehat{\omega}_k) \widehat{u}_{p,k}^-
		     \widehat{S}_-^k \right) \, ,
\label{commute}
\end{equation}
where equation (\ref{powereq}) has been used. The order of the operators 
$\widehat{u}_{p,k}^\pm$ and the frequency operator of the unperturbed system, 
$\widehat{\omega}_k$, is immaterial since they depend on $\widehat S_z$ only. 
Obviously, there is no term depending on 
$\widehat{S}_z$ alone; hence, for (\ref{integrate1}) to hold one must have
\begin{equation}
{\cal H}_p ( \widehat{S}_z) = R_{p,0}(\widehat{S}_z) \, .
\label{lowestorder}
\end{equation}
Furthermore, the operators $\widehat{u}_{p,k}^\pm$ follow from comparing 
(\ref{expandr}) and (\ref{commute}):  
\begin{equation}
u_p(\widehat{S}_\pm,\widehat{S}_z) = u_{p,0}(\widehat{S}_z)
		  +\sum_{k=1}^\infty
		  \left( \widehat{S}_+^k \frac{\widehat{R}_{p,k}^+}{i 
					 \widehat{\omega}_k}
		     + \frac{\widehat{R}_{p,k}^-}{-i\widehat{\omega}_k}
					  \widehat{S}_-^k
		  \right) \, ,
\label{resultate}
\end{equation}
where the inverse of the operator $\widehat{\omega}_k $ is well-defined only  
if the operator $\widehat{\omega}_k $ does not have a zero eigenvalue; 
this, however, is guaranteed by the assumption that the unperturbed 
Hamiltonian has no degenerate eigenvalues. 
Therefore, an operator $\widehat{u}_p$ solving Eq. (\ref{integrate1}) has 
been found in terms of $\widehat{R}_p$ and $\widehat{\omega}_k$. 
Its `diagonal' part $u_{p,0}(\widehat{S}_z)$ (corresponding to a phase shift 
when acting on eigenstates of $\widehat S_z$) remains undetermined---for 
simplicity, it will be chosen equal to zero for all $p$. 
As a result, the possibility to solve the hierarchy of equations 
(\ref{hierarchy1}) has been established. The similarity of Eq. 
(\ref{resultate}) with the classical formula, Eq. (\ref{wi1dim}), is striking.

It is important to note that the expression (\ref{resultate}) for the 
generator 
$\widehat{u}_p$ exhibits factors inversely proportional to the frequency 
operators $\omega_k (\widehat{S}_z)$, as is familiar from classical 
perturbation theory (cf. Appendix B). As can be read off from the structure 
of the hierarchy (\ref{hierarchy1}), the operators $\widehat{R}_p$ are 
linear in $\widehat{u}_{p-1}$: thus, they contain terms proportional 
to $1/\widehat \omega_k^{p-1}$. Solving  
(\ref{resultate}) for the generating operator of $p$th order, 
$\widehat{u}_p$, one finds it to be proportional to the inverse of the $p$th 
power of frequency operators. The structurally 
similar classical expansion is bound to converge to a well-defined canonical 
transformation since all classical systems with one degree of freedom are 
integrable. By analogy, it is expected that the indivdual generating 
operators $\widehat{u}_p$ will add up to a sensible expression for the 
unitary transformation $\widehat{\cal U}_\varepsilon$. 
If classical systems with two degrees of freedom are studied, the 
impact of a perturbation is known to be disastrous in most cases. 
\subsection{Two degrees of freedom}
For a nearly-integrable quantum system with two degrees of freedom
\begin{equation}
H (\widehat{ \bf S}_{1}, \widehat{   \bf S}_{2}, \varepsilon) 
	= H_{0} (\widehat  S_{1z}, \widehat  S_{2z}) 
	  + \widetilde H ( \widehat{ \bf S}_{1}, \widehat{ \bf S}_{2}, 
	   \varepsilon) \, ,
\label{2nearly}
\end{equation}
the transformed Hamiltonian operator should depend on two new spin 
components only, $\widehat {\cal S}_{1z}$ and $\widehat {\cal S}_{2z}$.
As before, the lowest order of the hierarchy (\ref{hierarchy1}) is easily 
solved by defining $\widehat{\cal H}_0$ to be identical with the unperturbed  
Hamiltonian, $\widehat H_0$. For $p\geq 1$, the equations to be solved read:
\begin{equation}
\frac{i}{\hbar}[H_0(\widehat{S}_{1z},\widehat{S}_{2z}), 
		 u_p(\widehat{S}_{1\pm},\widehat{S}_{2\pm},
				     \widehat{S}_{1z}, \widehat{S}_{2z})] 
       = R_p(\widehat{S}_{1\pm},\widehat{S}_{2\pm},
				   \widehat{S}_{1z},\widehat{S}_{2z}) 
	      - {\cal H}_p (\widehat{S}_{1z}, \widehat{S}_{2z}) \, .
\label{integrate2}
\end{equation}
Using expansions for the operators $\widehat R_p$ and $\widehat u_p$ 
(cf. Appendix C) in analogy to (\ref{expandr}) and (\ref{expandu}), 
respectively, one finds for the generator   
\begin{eqnarray}
\widehat{u}_p (\widehat{\bf S}_\pm,\widehat{\bf S}_z) 
& = & \sum_{k,\ell=0 \atop {(k,\ell) \neq(0,0)}}^\infty
\left( \widehat{S}_{1+}^k \widehat{S}_{2+}^\ell 
	   \frac{R_{p,k\ell}^+ (\widehat{\bf S}_z)} 
      {i (\omega_{k}^{(1)} (\widehat{S}_{1z},\widehat{S}_{2z} + \ell \hbar) 
       + \omega_{\ell}^{(2)} (\widehat{S}_{1z}, \widehat{S}_{2z}))} 
       \right.  \nonumber \\
&   &  \left. + \widehat{S}_{1+}^k
\frac{R_{p,k\ell}^\pm (\widehat{\bf S}_z)}{i (\omega_{k}^{(1)}
		      (\widehat{\bf S}_z)-\omega_{\ell}^{(2)} 
		   (\widehat{\bf S}_z))} \widehat{S}_{2-}^\ell + \mbox{h.c.} 
			   \right)\, .
\label{twodimu}
\end{eqnarray}
Again, the generator $\widehat u_{p}$ is determined up to an arbitrary 
function of $ \widehat{\bf  S}_{z}$ only which is chosen equal to zero. 
When expressing the denominators in terms of the unperturbed Hamiltonian,
it becomes obvious that both spins enter symmetrically in (\ref{twodimu}): 
the first one reads
\begin{equation}
\widehat{\Delta}^+_{k\ell}
	  = \frac{1}{\hbar} 
	  \left( H_0 ( \widehat{S}_{1z} + k \hbar, 
				\widehat{S}_{2z} + \ell \hbar)  
	  -  H_0 ( \widehat{S}_{1z}, \widehat{S}_{2z} )  \right) \, ,
\label{deltaplus}
\end{equation}
and the second one is equal to 
\begin{equation}
\widehat{\Delta}^-_{k\ell}
	  = \frac{1}{\hbar} 
	  \left( H_0 ( \widehat{S}_{1z} + k \hbar, \widehat{S}_{2z} )  
	  -  H_0 ( \widehat{S}_{1z}, \widehat{S}_{2z} + \ell \hbar) 
	    \right) \, .
\label{deltaminus}
\end{equation}
Eq. (\ref{twodimu}) belongs to the main results of the present paper. 
For systems with two (or more) degrees of freedom the generators contain 
terms proportional to the inverse of {\em linear combinations\/} of frequency 
operators. The expansion (\ref{twodimu}) makes sense only if the denominators 
$\widehat{\Delta}^\pm_{k\ell}$ do not have a zero eigenvalue, which will be 
assumed from now on. When dealing with the convergence of the perturbation 
series, it will be necessary to proceed in two steps: first, the existence of 
each generator 
$\widehat{u}_p$ has to be shown, and subsequently it has to be checked that 
the sum of all $\widehat{u}_p$ leads to a sensible result. 
\section{Examples}
The pertubation theory as developed in the previous 
section is applied here to explicit examples. Systems with one and two 
degrees of freedom will be studied in order to obtain more detailed insight 
into the structure of the denominators.
\subsection{One degree of freedom}
The Hamiltonian to be studied is defined as follows 
\begin{eqnarray}
H (\widehat{\bf S}, \varepsilon)  
& = & \alpha \widehat S_{z} 
	   + \frac{1}{2} \beta \widehat S_{z}^{2} 
	   + \frac{\varepsilon}{2} \widehat S_{x}^{2} \nonumber\\ 
& = & \alpha \widehat S_{z} + \frac{1}{2} \beta \widehat S_{z}^{2} 
     + \frac{\varepsilon}{4} ({\bf \widehat S}^{2} - \widehat S_{z}^{2})
     + \frac{\varepsilon}{8} (\widehat S_{+}^{2} + \widehat S_{-}^{2}) \, ,
\label{explicitH1}
\end{eqnarray}
the nonzero constants $\alpha$ and $\beta$ being real numbers, and 
the quantum integrable part, 
$\widehat H_{0} = \alpha \widehat S_{z} + \beta \widehat S_{z}^{2}/2$,
is assumed to have no degenerate eigenvalue.
The perturbation 
$\widetilde H ( \widehat{\bf S},\varepsilon) 
		  = \varepsilon \widehat S_{x}^{2} /2 $
is particularly simple because it consists of just one single term being 
proportional to $\varepsilon$. 

It is straightforward to calculate the 
frequency operators according to (\ref{qmfreqpower}):
\begin{equation}
\widehat \omega_{k} 
      = k (\alpha + \beta \widehat S_{z}) + \frac{1}{2} \beta k^2 \hbar \, , 
	  \quad k=1,2, \ldots 
\label{onedimfreq}
\end{equation}
The first term on the right-hand-side defines a function $\omega_k$ 
of the operator $\widehat S_{z}$ which coincides with the expression for 
the classical frequency: 
$\omega_{k} (S_{z}) = k \omega (S_{z}) 
		    = k (\alpha + \frac{1}{2} \beta S_{z})$.
Due to the discretization of the derivative an additional term proportional 
to $\hbar$ shows up in (\ref{onedimfreq}).

According to (\ref{lowestorder}) the new first-order Hamiltonian is equal 
to the diagonal part of the perturbation:
\begin{equation}
{\cal  H}_1 (\widehat S_{z}) 
	= \frac{1}{4} ({\bf \widehat{ S}}^{2} - \widehat S_{z}^{2}) \, ,
\label{1orderdiag}
\end{equation}
as can be read off from Eq. (\ref{explicitH1}). When expanding the 
off-diagonal part of the perturbation the only nonvanishing coefficients are 
found to be
\begin{equation}
R^{\pm}_{1,2} (\widehat S_{z}) = \frac{1}{8} \, ,
\label{1offdiag}
\end{equation}
leading to the following expression for the first-order generator 
\begin{equation}
\widehat u_1 = \frac{1}{8} \left( \widehat S_{+}^{2} \frac{1}{i \widehat 
								 \omega_{2}} 
		 + \frac{1}{-i \widehat \omega_{2}} \widehat S_{-}^{2} \right)
		\, ,
\label{1ordergen}
\end{equation}
with $\widehat \omega_2$ from (\ref{onedimfreq}).
Since the energy eigenvalues of the Hamiltonian  $\widehat H_0$ in 
(\ref{explicitH1}) have been assumed not to be degenerate, the expectation 
value of the denominators in Eq. (\ref{1ordergen}) will not have zero 
eigenvalues. Consequently, there is no principal    
obstacle which would destroy the convergence of the perturbation series 
for the unitary transformation $\widehat{\cal U}$, similar to classical 
mechanics. 
\subsection{Two degrees of freedom}
Consider a two-spin Hamiltonian (\ref{2nearly}) with
\begin{equation}
H (\widehat{\bf S}_{1}, \widehat{\bf S}_{2}, \varepsilon) 
      =  f_1 (\widehat S_{1z}) + f_{2} (\widehat S_{2z}) 
		       + \gamma \widehat S_{1z} \widehat S_{2z}
	 + \widetilde H (\widehat{\bf S}_{1}, 
			 \widehat{\bf S}_{2}, \varepsilon) \, ,
\label{2dimexample}
\end{equation}
where 
\begin{equation}
f_{j} (x) = \alpha_{j} x + \frac{1}{2} \beta_{j} x^{2} \, , 
	     \qquad j = 1, 2 \, .
\label{functionf}
\end{equation}
The quantum integrable Hamiltonian consisting of the first three terms  
on the right-hand-side of Eq. (\ref{2dimexample}) is quadratic in the 
$z$ components of the spin. With a view to the structure of the denominators 
present in the generating operator---which are completely determined by the 
frequencies of the unperturbed system as seen from Eq. (\ref{twodimu})---the 
form of the perturbation $ \widetilde H (\widehat{\bf S}_{1}, 
\widehat{\bf S}_{2}, \varepsilon)$ must not be specified 
in detail. On evaluating (\ref{Nfreq}) one finds
\begin{equation}
\widehat\omega_{k}^{(1)} 
	 =  k (\alpha_{1} + \beta_{1} \widehat S_{1z} 
			  + \gamma \widehat S_{2z}) 
	   + \frac{1}{2} \beta_{1} k^{2} \hbar \, ,
\label{qmfrequencies}
\end{equation}
and $\widehat\omega_{k}^{(2)}$ follows from  exchanging the indices  
$(1 \leftrightarrow 2)$. The first term is closely related to the 
classical frequency, 
\begin{equation}
\omega_{k}^{(1)} (S_{1z},S_{2z}) 
     = k \omega^{(1)} (S_{1z}, S_{2z}) 
     = k \frac{\partial H_0 (S_{1z}, S_{2z}) }{\partial S_{1z}} 
     = k (\alpha_{1} + \beta_{1} S_{1z} + \gamma S_{2z}) \, ,
\label{clfrequencies}
\end{equation}
and analogously for $\omega_{k}^{(2)} (S_{1z},S_{2z})$. In formula 
(\ref{twodimu}), the expansion of the generator $\widehat u_p$ has been
given in general terms. Focus now on the second term of its right-hand-side
containing the difference of the frequency operators in the denominator. 
Its matrix elements in the product basis $| m_{1}, m_{2} \rangle$, consisting 
of eigenstates of the operators $\widehat S_{jz}$, are given by
\begin{equation}
\langle m_1 , m_2 | \widehat S_{1+}^k 
	   \frac{\widehat R^+_{p,kl}}{\widehat \Delta^-_{k \ell}}
		\widehat S_{2-}^\ell | m_1^\prime , m_2^\prime \rangle 
= 
\langle m_1^\prime , m_2 | 
	 \frac{\widehat R^+_{p,k\ell}}{\widehat \Delta^-_{k \ell}}
			 | m_1^\prime , m_2 \rangle 
	     \delta_{m_1 m_1^\prime + k} \, \delta_{m_2 m_2^\prime-\ell} \, ,
\label{matrixelements}
\end{equation}
where, as before, $ \widehat\Delta^-_{k\ell} 
	 = \widehat\omega_{k}^{(1)} - \widehat\omega_{\ell}^{(2)}$,
which in the present case reads explicitly 
\begin{equation}
\widehat\Delta^-_{k\ell} 
      = \left( k (\alpha_1 + \beta_1 \widehat S_{1z} 
		   + \gamma \widehat S_{2z} ) 
	    - \ell (\alpha_{2} + \beta_{2} \widehat S_{2z} 
			  + \gamma \widehat S_{1z}) \right) 
	    + \frac{\hbar}{2} (\beta_1 k^{2} - \beta_{2} \ell^{2})
		      \, .
\label{denominatorop}
\end{equation}
Since both operators, $\widehat R^+_{p,k\ell}$ and $\widehat 
\Delta^-_{k \ell}$, depend on the $z$ components only, the expectation value 
of their ratio, $\langle \widehat R / \widehat \Delta \rangle$, is equal to 
the ratio of their expectation value, $\langle \widehat R \rangle / \langle 
\widehat \Delta \rangle$. Consequently, the convergence of the power series 
defining the first-order generator for the system (\ref{2dimexample}), will 
be controlled by matrix elements of operators such as 
$\widehat\Delta^-_{k\ell}$ in Eq. (\ref{denominatorop}). They are made up
from two contributions with different origin. The operator in large round 
brackets is recognized as the 
difference of the frequency operators which one would obtain if in the 
classical formula (\ref{clfrequencies}) one were to replace the variables 
$S_{jz}$ by their operator equivalents, $\widehat S_{jz}$. 
The possibility to approximate arbitrarily well the ratio of the 
frequencies $\omega^{(2)}/\omega^{(1)}$ for a given torus (i.e. fixed values 
of $S_{1z}$ and $S_{2z}$) by $k/\ell$ leads to the problem of small 
denominators in classical mechanics (cf. \cite{thirring88}).

Then, the difference $\widehat\Delta^-_{k \ell}$ contains a term 
proportional to Planck's constant: the occurrence of a term {\em linear\/} 
in $\hbar$ follows from studying a Hamiltonian $\widehat H$ being quadratic 
in the $z$ components of the spins. 
Polynomials of degree $r$ for the Hamiltonian lead to terms of at most 
$\hbar^{r-1}$, etc. The presence of such terms has been noted already by 
Robnik \cite{robnik86} in a semiclassical treatment of particle systems. 
He suggested that these additional terms will have the effect to move 
the system away from the classical resonance, however small the value of 
$\hbar$. But looking at the {\em exact\/} expression (\ref{denominatorop}) for 
a fixed value of Planck's constant, $\hbar = 1$, say, one can also conceive 
it as a new type of resonance condition which is no longer {\em linear\/} in 
the indices $k$ and $\ell$: a quantum-mechanical {\em nonlinear\/} resonance 
condition arises which has to be studied in its own right for all possible
representations. 

The relevant matrix elements of the operator $\widehat\Delta^-_{k \ell}$ 
(cf. (\ref{matrixelements})) read explicitly
\begin{equation}
\langle m_1-k, m_2^\prime -\ell | \widehat\Delta^-_{k \ell} 
		 | m_1-k, m_2^\prime- \ell \rangle 
	= k (\alpha_1 + \beta_{1} \hbar m_{1} + \gamma \hbar m_2^\prime) 
	  - \ell (\alpha_{2} + \beta_{2} \hbar m_{2}^\prime 
			     + \gamma \hbar m_1)  
	  - \frac{\hbar}{2} (\beta_{1} k^{2} - \beta_{2} \ell^{2}) \, , 
\label{quadraticform}
\end{equation}
(note the sign change of the last term relative to
(\ref{denominatorop})).
In each individual representation, there is an important difference compared  
to classical mechanics. The range of the 
numbers $k$ and $\ell$ is not all the integers but it varies over a finite
subset of them only, which is determined by the spin length $s$. This is 
characteristic of finite-dimensional representations of spin systems while  
it does not occur for particle systems. The matrix elements of the other 
operators in Eq. (\ref{twodimu}) lead to expressions of the same structure.

There are two ways to study the expression (\ref{quadraticform}). First, 
consider it as a quantum-mechanical statement in its own right. This means to 
give $\hbar$ 
its actual value and to let vary both pairs $(m_{1}, m_{2}^\prime)$ and 
$(k, \ell)$ 
over the finite range determined by the value of $s$. Within each $(2 
s+1)$-dimensional representation of the spin algebra, the modulus of 
$\langle \widehat \Delta^-_{k \ell} \rangle $ will have some positive nonzero 
minimal value, $\Delta (s)$. Consequently, for sufficiently small values of 
$\varepsilon$, the expansion of the generator $\hat u_{p}$ in (\ref{upower}) 
is expected to converge because the terms $(\varepsilon / \Delta 
(s))^{p}$ can be kept small. For spins of length $s$ the perturbed 
Hamiltonian $\widehat H_{\varepsilon}$ is 
given by a finite-dimensional hermitian matrix which guarantees the 
existence  of  a unitary transformation diagonalizing it. Therefore, the 
generic convergence of the construction should not come as a surprise (for 
possible technical difficulties, cf. \cite{jauslin+95}). For the present 
concept of integrability, however, it is essential that not one {\em 
specific\/} representation is studied but that the statements hold in an 
{\em algebraic\/} sense, or, equivalently, in {\em all\/} representations. The 
implications of this requirement will be made explicit now. 

Imagine to have constructed the diagonalizing transformations in 
all representations. This set of matrices may either be the representations 
of one {\em single\/} algebraically defined operator $\widehat{\cal 
U}_{\varepsilon}$ 
or not. In the latter case, the unitary transformations in the individual 
representations will not ``converge'' towards one specific unitary 
operator for large values of $s$ (since the value of $\hbar$ is kept fixed, 
this procedure is not identical to the classical limit to be discussed 
momentarily). In addition, for larger values of $s$ the number $\Delta (s)$ 
takes on smaller and smaller values which requires the expansion parameter 
to be correspondigly smaller. If there is no {\em finite\/} value of 
$\varepsilon$ such that the perturbation series converges for all allowed 
values of $s$, i.e. algebraically, the perturbation $\widetilde H 
(\widehat{\bf S}, \varepsilon)$ renders the system non-integrable. A strong 
decrease of 
the smallest  occurring denominator $\Delta (s)$ as a function of $s$ 
has been observed numerically. If, on the other hand, the summation of the 
perturbation series for all $s$ does not require vanishing $\varepsilon$, 
one may expect an algebraic diagonalizing operator to exist; hence, the 
perturbation can be absorbed into an appropriate redefinition of the action 
operators. In classical mechanics, this convergence is known for finite 
$\varepsilon$ is known to result from appropriately decreasing numerators. 

Turn now to the second way to study (\ref{quadraticform}): in the 
semiclassical limit with $\hbar \sim 1/s$ and $s \rightarrow 
\infty$, one should recover the behaviour of the classical frequency 
denominators from Eq. (\ref{quadraticform}). Numerical tests make it 
plausible that, indeed, the value of $\Delta (s)$ decrease monotonically 
when larger values of $k$ and $\ell$ provide better and better approximations 
of the classical frequency ratio (the values of $\hbar m_1$ and $\hbar 
m_{2}^\prime$
are kept fixed here defining thus a specific ``torus''). An analytic study is 
not straightforward, even for the quadratically modified resonance condition  
associated with (\ref{quadraticform}): the behaviour of the last 
term is not easily controlled since $\hbar k^{2}$ and $\hbar \ell^{2}$
may have values of the order of $s$.

\section{Discussion}
\subsection{Summary and outlook}
For the sake of clarity, the main points of the development given in the 
previous pages are briefly summarized. Starting from the algebraic skeleton 
necessary to describe quantum (spin) systems, arguments for the existence of 
quantum-mechanical action and 
frequency operators in systems with a single degree of freedom have been 
put forward. When applying these concepts to systems with more degrees 
of freedom they lend themselves to a definition of quantum integrable 
systems. An algebraic version of perturbation theory of such systems 
has been formulated capable to absorb the effect of an added disturbance. 
Since this formulation respects (and exploits) the underlying Lie-algebraic 
structure, a tight analogy to classical Lie transforms is achieved. As a 
natural consequence, operator-valued analogs of the classical frequency 
denominators show up when constructing a unitary transformation which would 
diagonalize the Hamiltonian algebraically. In this way, the 
quantum-mechanical locus of the convergence problems known to exist in 
classical 
mechanics has been found, at least on a formal level. In an appropriate 
sense, the quantum-mechanical expressions for the denominators generalize 
the classical ones. The condition for a classical resonance to occur is, 
effectively, equivalent to a {\em linear\/} relation among the classical 
frequencies (associated with a ``plane'' in action space) whereas, in a 
given representation, the relevant quantum-mechanical relation turns out 
to be {\em nonlinear\/} (defining ``ellipsoids'' or other curved surfaces). 
Such structures have been observed in semiclassical treatments, for example, 
in the work by Berry and Tabor \cite{berry+77} who derive energy-level 
statistics for classically integrable quantum systems in the limit of small 
values for $\hbar$.

When scrutinizing the notion of quantum integrability the ultimate goal is 
to provide evidence for both, its usefulness and its consistency. Rigorously 
testable predictions of 
specific properties must result for quantum integrable (and nonintegrable) 
systems, just as the phase-space foliation of classically integrable  
systems flows from its classical counterpart. In our view, the major 
achievement of the present approach is to have boiled 
down the question of quantum integrability (and, {\em a fortiori\/}, of 
quantum 
chaos) to a {\em technical\/} problem, namely to investigate the (convergence) 
properties of a perturbation series. Due to the {\em nonlinearities\/} in the 
quantum-mechanical expressions just mentioned, the behaviour of the 
perturbation series is, from the present point of view, not obvious. 
In principle, there is no need 
for quantum systems to {\em exactly\/} parallel the properties of classically 
(non-) integrable systems: the underlying unifying structure is on an {\em 
algebraic\/} level but with the algebras being realized in different spaces, 
in phase space and in Hilbert space, respectively. 
In the semiclassical limit, however, one expects that the statements 
associated with the two realizations become comparable if appropriate tools  
such as Wigner functions are employed. 
The present formulation now calls for a detailed parallel investigation of 
the effects which a perturbation has on a classically integrable system and 
on its quantum counterpart. 
 
Here is a brief sketch of a scenario being plausible in view of the present 
results which, however, has to be substantiated by further investigations. 
Focus on the fully quantum-mechanical regime, that is, fix the value of 
$\hbar$ and consider the behaviour of the perturbation series in 
representations of variable dimension, $2 s+1$. It is expected that the 
``visibility'' of the perturbation will increase for dimensions with larger 
values of $s$. 
A vivid illustration of this effect has been reported for a system consisting 
of two coupled spins where a ``regular quantum web'' is gradually torn apart 
by a perturbation of increasing strength \cite{srivastava+90}. Related 
phenomena have been observed in particle systems by Bohigas, Tomsovic, and 
Ullmo \cite{bohigas+93}. In a study of the spin-boson model the observed 
``resonance phenomena'' can be attributed to the presence of small 
denominators becoming more or less effective in representations of different 
dimensions \cite{cibils+94}. An analytic 
study of these numerically observed effects in the vein of the present 
paper is under way \cite{mueller+95}. In general, ``quantum resonances'' 
in a nonintegrable quantum system due to ever smaller denominators will 
become more and more pronounced if larger representations of the spin algebra 
are considered. The infinite-dimensional representation (not to be confused 
with the semiclassical limit) contains the most detailed structure, possibly 
leading to an actual divergence of the perturbation series. 
\subsection{Related work}
There is a large number of joints between earlier work and the present one 
which so far have been alluded to only occasionally. It is useful to dwell 
on them for a moment, to draw parallels and to point out differences. 

The introduction of frequency operators (in the present terminology) provides 
a link to the most advanced papers of the ``old quantum mechanics.''  
In 1925, Dirac \cite{dirac25} defined ``q-numbers'' which would 
fulfill the fundamental commutation relations of position and momentum. 
Basic {\em algebraic\/} consequences are derived, partly by 
exploiting the analogy to the classical Poisson brackets. One section of 
the paper is devoted to ``multiply periodic systems'' characterized by the 
existence of ``uniformizing variables'' (which nowadays would be called 
action and angle operators) such that the Hamiltonian depends on the action 
operators only (cf. Eq. (\ref{Htrf})). In the terminology of the present work 
multiply periodic systems 
are thus recognized as quantum integrable ones. Assuming the action-angle 
operators to fulfill the same commutation relations as momentum and 
position---which is untenable due to the defectiveness of a phase operator 
\cite{newton80}---Dirac introduced two different types of frequencies. 
The first one is supposed to govern the time evolution of the phase 
operator itself whereas the second one is associated with the time derivative 
of the {\em exponentiated\/} phase operator. It turns out that the latter 
frequency operator is equivalent to the expression given in 
(\ref{qmfreq}). Dirac also mentions its two possible versions reflecting the 
freedom of ordering in (\ref{qmeqmo}). For spin systems considered here, 
the operator $\widehat  S_{+}$ plays a r\^ole similar to the exponentiated 
angle operator.

Wentzel \cite{wentzel26} continued Dirac's work by
establishing a link between Heisenberg's matrix mechanics and the 
q-numbers. In particular, he wrote down expressions for quantum-mechanical 
frequencies similar to (\ref{qmfreq}) being applicable to multiply periodic 
systems only. However, this line of reasoning seems to have come to an abrupt 
end by the advent of wave mechanics in 1926: the serious shortcoming  
of the earlier quantum theory to allow for quantization of multiply periodic 
systems only was removed by Schr\"odinger's equation  
which is easily written down for particles subjected to arbitrary potentials.

Similarly, strong formal affinities to Heisenberg's formulation of 
perturbation theory \cite{heisenberg26} exist. This is easily 
understood if one recalls that before the introduction of Schr\"odinger's 
equation classical mechanics served much more than later as Ariadne's 
thread for quantum-mechanical developments. Still, Heisenberg's approach to 
treat perturbations and the widely used perturbation scheme of 
Rayleigh-Schr\"odinger \cite{rayleigh+53} differ conceptually in an important 
way from the present one. 
Traditionally, {\em any\/} quantum system may serve as a starting point 
for perturbation theory, the only assumption being that its 
(necessarily periodic) solutions are known. In the approach developed here, 
the unperturbed system is assumed to be quantum integrable, 
hence it presumably stems from a {\em restricted\/} class of systems, and the 
effect of the perturbation is to remove it from this class. 
In order to prove the KAM-theorem, classical perturbation theory 
is developed relative to {\em integrable\/} systems: the fate of a torus 
under the perturbation is investigated, not its effect on an arbitrary 
periodic solution---which might happen to be an isolated periodic orbit of a 
nonintegrable system. KAM-theory would not apply at all.
It is hoped that by systematically moving away from 
{\em integrable\/} quantum systems generic properties will 
be seen to emerge for nonintegrable ones.

As for a close-up of the relation between classical tori and Heisenberg's 
matrix mechanics which fits well into the scheme developed here, see the 
work by Greenberg, Klein and Li \cite{greenberg+95}. 

Next, there are close ties to Birkhoff normal forms \cite{lichtenberg+83} and 
to the so-called ``algebraic quantization'' reviewed in \cite{fried+88}. The 
idea is to quantize integrable approximations to classically nonintegrable 
systems.  Crehan \cite{crehan90} worked out quantum-mechanical normal  
forms in the spirit of Birkhoff and Gustavson, starting from the classical 
Hamiltonian function in the neighbourhood of a point of stable equilibrium.
Impressive agreement between exact quantum-mechanical solutions and the 
approximate ones has been obtained \cite{uzer+83,farelly86}. In contrast to 
these works, however, no recourse has been made to classical mechanics in the 
present approach. 

Concerning the presence of the quantum equivalents of the small 
denominators, their occurrence in the semiclassical limit has been noted 
already by Robnik \cite{robnik86} (and by Graffi, cf. the introduction of 
\cite{bellissard+90}). Robnik pointed out that for small 
values of Planck's constant the classical denominators will be modified by 
terms of 
the order $\hbar$. It has been conjectured that in this way the resonance 
conditions are destroyed which govern the fate of an individual classical 
torus, that is, its survival or destruction. It is expected that in view of 
the {\em exact\/} expression (\ref{twodimu}) the effect of the denominators 
can be studied in more detail. 

Also, the understanding of clashes between the behaviour of classical 
and quantum perturbation expansions may benefit from the present 
approach. It has been noted by Eckhardt \cite{eckhardt86} for a 
one-dimensional particle system that the convergence of a classical 
perturbation series is not automatically reflected in the convergence of 
the corresponding quantum perturbation series as follows from a simple 
scaling argument.

In order to deal with issues of perturbation series for quantum systems 
a new setting has been provided by Bellissard and Vittot 
\cite{bellissard+90} whose work is based on the concept of non-commutative 
geometry. Their work is focussed on Lie-transform techniques for 
quantum-mechanical particle 
systems with respect to the classical limit. The aspect of integrability, 
seemingly, has not been addressed in detail. It is particularly 
important that estimates of the Nekhoroshev-type become available which allow 
one to control the approximations for energy eigenvalues extremely well. 
For systems with one degree of freedom, the convergence of the perturbation 
series has been proven (cf. \cite{bambusi95}). At present, the theory has 
been worked out for 
particle systems only while the transfer to spin systems is in preparation 
\cite{private}. A common feature of many works in this area (cf. also 
\cite{graffi+87} is that the unperturbed systems considered are collections 
of harmonic oscillators. Having frequency operators at one's diposal, this 
restriction seems to be no longer necessary. 

Finally, a review of the use of Lie-transform techniques and KAM-like results 
in quantum mechanics has been given by Jauslin \cite{jauslin93} dealing to a
large extent with externally driven systems.
\section*{A: The Commutators $[f(\widehat{S}_z),\widehat{S}_\pm^k]$}
\label{appA}
It will be shown by using the commutation relations of the spin operators 
that
\begin{equation}
[f(\widehat{S}_z),\widehat{S}_+^k] 
    = \widehat{S}_+^k \left( f(\widehat{S}_z+k\hbar) - f(\widehat{S}_z) \right)
    = \widehat{S}_+^k \Big( \exp (k \hbar \partial_z ) - 1 \Big) 
	  f(\widehat{S}_z) \, , 
\label{spkcom}
\end{equation}
where $\partial_z$ is a shorthand for a (formal) derivative with 
respect to $\widehat{S}_z$. The function $f(x)$ is assumed to have an 
expansion as a power series in $x$.

First, it is proved by induction that
\begin{equation}
[\widehat{S}_z^n,\widehat{S}_+]
 = \widehat{S}_+  \Big( \exp(\hbar\partial_z) - 1 \Big) \widehat{S}_z^n
\, . \label{sznsp}
\end{equation}
This relation holds for $n=1$,
\begin{equation}
[\widehat{S}_z,\widehat{S}_+]
  = \widehat{S}_+ \Big( \exp(\hbar\partial_z) - 1 \Big) \widehat{S}_z
  = \widehat{S}_+ (\widehat{S}_z+\hbar-\widehat{S}_z)=\hbar\widehat{S}_+ \, .
\end{equation}
Eq. (\ref{sznsp}) is valid for $n+1$ if assumed to hold for $n$:
\begin{eqnarray}
[\widehat{S}_z^{n+1},\widehat{S}_+] & = & [\widehat{S}_z,\widehat{S}_+] 
      \widehat{S}_z^n+\widehat{S}_z[\widehat{S}_z^n,\widehat{S}_+] 
				       \nonumber \\
& = & 
\widehat{S}_+\left(\hbar\widehat{S}_z^n+(\widehat{S}_z+\hbar)(\exp( 
	      \hbar\partial_z)-1)\widehat{S}_z^n \right) \nonumber \\
& = & \widehat{S}_+\Big(\exp(\hbar\partial_z)-1\Big)\widehat{S}_z^{n+1} \, ,
\end{eqnarray}
since $(\widehat{S}_z+\hbar)\exp(\hbar\partial_z)=\exp(\hbar\partial_z) 
			    \widehat{S}_z$.
Using $f(\widehat{S}_z)=\sum_{n=0}^\infty f_n\widehat{S}_z^n$ implies:
\begin{eqnarray}
[f(\widehat{S}_z),\widehat{S}_+] & = & \sum_{n=0}^\infty f_n[\widehat{S}_z^n,
			     \widehat{S}_+]
=\widehat{S}_+\Big(\exp(\hbar\partial_z)-1\Big)\sum_{n=0}^\infty f_n 
			       \widehat{S}_z^n \nonumber \\
& = & \widehat{S}_+\Big(\exp(\hbar\partial_z)-1\Big)f(\widehat{S}_z)\, .
\label{lowestk}
\end{eqnarray}
Thus, Eq. (\ref{spkcom}) holds for $k=1$, as is necessary for a proof by 
induction on $k$. The step from $k$ to $k+1$ reads
\begin{eqnarray}
[f(\widehat{S}_z),\widehat{S}_+^{k+1}] & = & 
\widehat{S}_+^k[f(\widehat{S}_z),\widehat{S}_+]+[f(\widehat{S}_z),
	 \widehat{S}_+^k]\widehat{S}_+ \nonumber \\
& = & \widehat{S}_+^{k+1}\Big(\left(\exp(\hbar\partial_z)-1\right)+
\left(\exp(k\hbar\partial_z)-1\right)\exp(\hbar\partial_z)\Big)
				      f(\widehat{S}_z) \nonumber \\
& = & \widehat{S}_+^{k+1}\Big(\exp\left((k+1)\hbar\partial_z\right)-1\Big)
f(\widehat{S}_z)\, ,
\end{eqnarray}
where, in the second line, Eq. (\ref{lowestk}) has been used in the form:
$f(\widehat{S}_z)\widehat{S}_+ = \widehat{S}_+\exp(\hbar\partial_z)f(
   \widehat{S}_z)$, concluding the proof of (\ref{spkcom}).

Relation (\ref{spkcom}) can also be written as
\begin{equation}  
f(\widehat{S}_z)\widehat{S}_+^k=\widehat{S}_+^kf(\widehat{S}_z+k\hbar)\, , 
			 \quad\mbox{or}\quad
f(\widehat{S}_z-k\hbar)\widehat{S}_+^k=\widehat{S}_+^kf(\widehat{S}_z)\, ,
\label{otherform}
\end{equation}
in words, the order of a function of $\widehat{S}_z$ only and the $k$th power 
of the operator $\widehat{S}_+$ may be exchanged if the argument of the 
function $f$ is shifted by an appropriate multiple of $\hbar$.
Similar relations are obtained for the step-down operator by taking the 
hermitian conjugate of Eqs. (\ref{spkcom}) and (\ref{otherform}).

A proof of Eq. (\ref{lowestk}) for {\em arbitrary\/} functions 
$f (\widehat S_z) $ is easily given if one exploits the properties of the 
ladder operators in explicit representations. The action of the step-up 
operator is in all representations of the form 
\begin{equation}
\widehat S_+ | m,s \rangle = c_m^s | m,s \rangle \, , 
	      c_m^s = \sqrt{s(s+1) - m(m+1)} \, .
\label{plusaction}
\end{equation}
Using $f (\widehat S_z) | m,s \rangle = f (m\hbar) | m,s \rangle$, one  
obtains for all $ | m,s \rangle $ that
\begin{eqnarray}
f (\widehat S_z) \widehat S_+ | m,s \rangle
    &=& c_m^s f ((m+1)\hbar) | m+1, s \rangle  \\
    &=& \widehat S_+ f ((m+1) \hbar) | m+1, s \rangle
	 = \widehat S_+ f ( \widehat S_+ \hbar) | m, s \rangle\, ,
\label{exchange}
\end{eqnarray}
implying that
\begin{equation}
f (\widehat S_z) \widehat S_+  = \widehat S_+  f (\widehat S_z+ \hbar) \, , 
\label{done}
\end{equation}
which is equivalent to (\ref{lowestk}). As before, induction on $k$ completes 
the proof of relation (\ref{spkcom}).
\section*{B: Classical perturbation theory}
Results of classical Lie-perturbation theory are collected, for easy 
comparison with the quantum-mechnical formulae an analogous notation is used.  
Consider a Hamiltonian  
$H({\bf S},\varepsilon) = H_0({\bf S}_z)+\widetilde{H}({\bf S},\varepsilon)$,
where $H_0({\bf S}_z)$ is integrable and $\widetilde{H}({\bf S},\varepsilon)$
is the perturbation. The goal is to remove the perturbation by introducing 
a new set of spin variables 
\begin{equation}
{\bf S}^\prime = \widehat T {\bf S} \, ,
\end{equation}
where the canonical transformation is implemented by the operator
\begin{equation}
\widehat T = \exp(-\{ u({\bf S},\varepsilon),\cdot\}) \, .
\end{equation}
The function $u({\bf S},\varepsilon)$ is assumed to have an expansion
in powers of $\varepsilon$, and $u({\bf S},\varepsilon=0)=0$.
When expanding the old and new Hamiltonian functions 
in powers of $\varepsilon$, the construction of the transformation 
$\widehat T$ leads to a hierarchy of equations
\begin{equation}
\{H_0({\bf S}_z),u_p({\bf S}_\pm,{\bf S}_z)\}
   = R_p({\bf S}_\pm,{\bf S}_z)-{\cal H}_p({\bf S}_z) \quad p= 0,1,2,\ldots 
   \, ,
\label{classhierarchy}
\end{equation}
which is the classical counterpart of (\ref{hierarchy1}). They can be 
solved by an appropriate choice of the functions $u_p$ and${\cal H}_p$. 

If in classical mechanics the expansion of functions $F(\widehat{\bf S})$ 
is written in analogy to (\ref{orderF1}) the solutions of 
(\ref{classhierarchy}) are
\begin{eqnarray}
{\cal H}_p(S_z) 
    & = & R_{p,0}(S_z)\, , \\
u_p(S_\pm,S_z) 
    & = & \sum_{k=1}^\infty \left( S_+^k 
    \frac{R_{p,k}(S_z)}{i\omega_k(S_z)} - \frac{R_{p,k}^ 
    \ast(S_z)}{i\omega_k(S_z)}S_-^k \right)\, . 
\label{wi1dim}
\end{eqnarray}
The expression for $u_p(S_\pm,S_z)$ can be simplified by using 
$\omega_k=k\omega$ from (\ref{clfrpower}).

Similarly, the results for a two-spin system assume the form  
of the expansion (\ref{orderF2}),
\begin{eqnarray}
{\cal H}_p({\bf S}_z) & = & R_{p,00}({\bf S}_z)\, , \\
u_p({\bf S}_\pm,{\bf S}_z) & = & \sum_{k,l=0\atop (k,l)\neq(0,0)}^\infty
\left( S_{1+}^kS_{2+}^l
\frac{R_{p,kl}^+({\bf S}_z)}{i(\omega^{(1)}_k({\bf S}_z)+\omega^{(2)}_l
({\bf S}_z))}\right.\nonumber \\
& & \left.
+S_{1+}^k\frac{R_{p,kl}^\pm({\bf S}_z)}
{i(\omega^{(1)}_k({\bf S}_z)-\omega^{(1)}_l({\bf S}_z))}
S_{2-}^l + \mbox{c.c.} \right) \, .
\end{eqnarray}
\section*{C: Expansion of a function $F(\widehat{S}_\pm,\widehat{S}_z)$}
A convention for the expansion of functions of spin operators is
established. For simplicity, the case of a single spin $\widehat{\bf S}$ is 
considered first.

Let $F(\widehat{S}_\pm,\widehat{S}_z)$ be expandable in powers of 
$\widehat{S}_+$ and 
$\widehat{S}_-$. Functions depending on $\widehat{S}_z$ only are denoted by 
curly symbols ${\cal F}(\widehat{S}_z)$. In general, $\widehat{F}$ will consist 
of a sum of terms, each of which is of the form
\begin{equation}  
\widehat{S}_+^{n_{1+}}\widehat{S}_-^{n_{1-}}{\cal F}_1(\widehat{S}_z) 
\widehat{S}_+^{n_{2+}}
\widehat{S}_-^{n_{2-}}{\cal F}_2(\widehat{S}_z) \cdots 
\label{term}
\end{equation}
The exponents $n_{j,\pm}$ are integers greater or equal to zero. Since 
the product $\widehat{S}_ + \widehat{S}_- = g(\widehat{S}_z)$ depends only on $
\widehat{S}_z$, one can write for arbitrary integers $n_\pm\ge 0$ that
\begin{equation}  
\widehat{S}_+^{n_+}\widehat{S}_-^{n_-} = \left\{
\begin{array}{ccc}
\widehat{S}_+^{n_+-n_-}{\cal F}_> (\widehat{S}_z) & & n_+>n_-\, , \\
{\cal F}_= (\widehat{S}_z) & \mbox{if} & n_+=n_-\, , \\
{\cal F}_<(\widehat{S}_z)\widehat{S}_-^{n_--n_+} & & n_+<n_-\, ,
\end{array}\right.
\end{equation}
by exchanging $g (\widehat{S}_z)$ with $\widehat{S}_\pm^k$ according to the 
rules (\ref{otherform}). Thus, expression (\ref{term}) can be cast into
one of these forms if now $n_\pm$ are defined as the total number of 
$\widehat{S}_\pm$ operators in (\ref{term}), $n_\pm=\sum_j n_{j\pm}$.

Collecting all terms in the expansion of $F(\widehat{S}_\pm,\widehat{S}_z)$ 
with the same number of $\widehat{S}_+$ or $\widehat{S}_-$, the operator 
$\widehat{F}$ can be written as
\begin{equation}  
F(\widehat{S}_\pm,\widehat{S}_z)
    = F_0(\widehat{S}_z) + \sum_{k=1}^\infty \left( \widehat{S}_+^k F_k^+(
						 \widehat{S}_z)
		       +F_k^- (\widehat{S}_z) \widehat{S}_-^k \right)\, ,
\label{orderF1}
\end{equation}
which will be considered as standard form. For the sake of generality, the 
index $k$ is assumed to run over all integers, while in a specific 
representation the sum in (\ref{orderF1}) will run over $2s+1$ values only. 
If $\widehat{F}$ is a hermitean operator then $\widehat{F}_0$ is hermitean 
and
\begin{equation}  
F_k^+(\widehat{S}_z) = \left(F_k^-(\widehat{S}_z)\right)^\dagger\, .
\label{adjoin}
\end{equation}

The generalization to two or more degrees of freedom is 
straightforward but the notation becomes more cumbersome.
First consider a function 
$F(\widehat{S}_{1\pm},\widehat{S}_{1z},\widehat{S}_{2\pm},\widehat{S}_{2z})$ 
depending on the components of two spins $\widehat{\bf S}_1$ and 
$\widehat{\bf S}_2$. As before, the 
function $F$ is assumed to have an expansion in powers of the spin 
components. Since the components of different spins commute, 
$\widehat{F}$ can be brought to the form:
\begin{eqnarray}
\lefteqn{F(\widehat{S}_{1\pm},\widehat{S}_{1z},\widehat{S}_{2\pm},
			 \widehat{S}_{2z}) 
  = F_{00}(\widehat{S}_{1z},\widehat{S}_{2z}) 
	   + \sum_{k,l=0 \atop{(k,l) \neq (0,0)}}^\infty 
 \left(\widehat{S}_{1+}^k \widehat{S}_{2+}^l F_{kl}^{++}(\widehat{S}_{1z},
		 \widehat{S}_{2z}) \right.} \nonumber \\
& & + \widehat{S}_{1+}^k F_{kl}^{+-}(\widehat{S}_{1z},\widehat{S}_{2z}) 
			     \widehat{S}_{2-}^l  
  \left. + \widehat{S}_{2+}^lF_{kl}^{-+} (\widehat{S}_{1z}, \widehat{S}_{2z}) 
			 \widehat{S}_{1-}^k
+F_{kl}^{--}( \widehat{S}_{1z}, \widehat{S}_{2z}) \widehat{S}_{1-}^k 
\widehat{S}_{2-}^l \right)\, .
\end{eqnarray}
If $\widehat{F}$ is a hermitean operator this expression simplifies to
\begin{equation}  
F(\widehat{\bf S}_1,\widehat{\bf S}_2) = F_{00}(\widehat{\bf S}_z)
	     +\sum_{k,l=0 \atop{(k,l) \neq(0,0)}}^\infty 
	     \left(\widehat{S}_{1+}^k \widehat{S}_{2+}^l F_{kl}^{++}(
					 \widehat{\bf S}_z)
	     +\widehat{S}_{1+}^k F_{kl}^{+-} (\widehat{\bf S}_z) 
		    \widehat{S}_{2-}^l + \mbox{h.c.} \right)\, ,
\label{orderF2}
\end{equation}
since, generalizing (\ref{adjoin})
\begin{equation}  
\widehat{F}_{kl}^{++}=(\widehat{F}_{kl}^{--})^\dagger,
\quad\widehat{F}_{kl}^{+-}=(\widehat{F}_{kl}^{-+})^\dagger\, .
\end{equation}

Functions depending on the components of more than two spins can be 
expanded similarly. One gets one term depending only on the $z$ 
components and $2^N$ terms which contain the explicit dependence on 
the operators $\widehat{S}_{j+}$ and $\widehat{S}_{j-}$.
A unique way of ordering would be, in extension of above, to throughout 
position the step-up operators $\widehat{S}_{j+}$ on the left and their 
adjoints 
on the right of the middle part which depends on the $z$ components only.

\begin{references}
%
\bibitem[*]{diplom}
	 This work is based on: 
	 Th. Gramespacher ({\sl Lie'sche St\"orungstheorie und die 
	 Integrabilit\"at quantenmechanischer Spinsysteme.})
	 Diplomarbeit (Universit\"at Basel, 1994).
\bibitem{lichtenberg+83}
	 A.\ J.\ Lichtenberg and M.\ A.\ Liebermann,  
	 {\sl Regular and Stochastic Motion.}
	 Springer, New York 1983.
\bibitem{giannoni+91}
	 {\sl Chaos and Quantum Physics.}
	 M.\ J.\ Giannoni, A.\ Voros, and J.\ Zinn-Justin (Eds.)
	 North-Holland, Amsterdam 1991.
\bibitem{weigert+95}
	 St.\ Weigert and G.\ M\"uller:
	 Chaos, Solitons \& Fractals {\bf 5}, 1419 (1995).
\bibitem{weigert92}
	 St.\ Weigert:
	 Physica D 56, 107 (1992).
\bibitem{falk51}
	 G.\ Falk:
	 Z.\ Phys.\ {\bf 130}, 51 (1951); {\bf 131}, 470 (1952). 
\bibitem{dirac25}
	 P.\ A.\ M.\ Dirac:
	 Proc. R. Soc. London, {\bf 110}, 561 (1926).
\bibitem{cary81}
	 J.\ R.\ Cary:
	 Phys. Rep. {\bf 79}, 131 (1981).
\bibitem{shavitt+80}
	 I.\ Shavitt and L.\ T.\ Redmon:
	 J. Chem. Phys. {\bf 73}, 5711 (1980).
\bibitem{sibert88}
	 E.\ L.\ Sibert III:
	 J.\ Chem.\ Phys.\ {\bf88} 4378 (1988).
\bibitem{thirring88}
	 W.\ Thirring:
	 {\sl Lehrbuch der Mathematischen Physik, I.}
	 Springer: New York, Heidelberg, Berlin 1988.
\bibitem{robnik86}
	 M.\ Robnik: 
	 J. Phys. A {\bf 19}, L841 (1986). 
\bibitem{jauslin+95}
	 H.\ R.\ Jauslin, M.\ Govin, and M.\ Cibils:
	 {\em Convergence of Iterative Methods in Pertubation Theory,\/}
	 in: {\em The Interplay betwenn Stochastics, Classics and Quanta.\/}
	 p. 151.
	 P.\ Garbaczewski, M.\ Wolf, and A.\ Weron (Eds.),
	 Lecture Notes in Physics, Springer 1995.
\bibitem{berry+77}
	 M.\ V.\ Berry and M.\ Tabor:
	 Proc.\ Roy.\ Soc.\ London {\bf A 356}, 375 (1977).  
\bibitem{srivastava+90}
	 N.\ Srivastava and G.\ M\"uller:
	 Z. Phys. B {\bf 81}, 137 (1990).
\bibitem{bohigas+93}
	 O.\ Bohigas, S.\ Tomsovic, and D.\ Ullmo:
	 Phys.\ Rep.\ {\bf 223}, 43 (1995).
\bibitem{cibils+94}
	 M.\ Cibils, Y.\ Cuche, and G.\ M\"uller:
	 Z.\ Phys.\ B {\bf 97}, 569 (1995).
\bibitem{mueller+95}
	 Private communication by G.\ M\"uller.
\bibitem{newton80}
	 R.\ G.\ Newton:
	 Ann. Phys. (N.Y.) {\bf 124}, 327 (1980).
\bibitem{wentzel26}
	 G.\ Wentzel:
	 Z. Phys. {\bf 37}, 80 (1926).
\bibitem{heisenberg26}
	 W.\ Heisenberg: 
	 Z. Phys. {\bf 33}, 879 (1925).
\bibitem{rayleigh+53}
	 See, e.g.: P.\ M.\ Morse and H.\ Feshbach:
	 {\sl Methods of Theoretical Physics} 
	 (McGraw-Hill, New York, 1953), Vol. 2, Chap. IX.
\bibitem{greenberg+95}
	  W.\ R.\ Greenberg, A.\ Klein, and C.-T.\ Li:
	  Phys.\ Rev.\ Lett.\ {\bf 75}, 1244 (1995).
\bibitem{fried+88}
	 L.\ E.\ Fried and G.\ S.\ Ezra:
	 J. Phys. Chem. {\bf 92}, 3144 (1988).
\bibitem{crehan90}
	 P.\ Crehan, J.\ Phys.\ A {\bf 23}, 5815 (1990). 
\bibitem{uzer+83}
	 T.\ Uzer, D.\ W.\ Noid, and R.\ A.\ Marcus:
	 J.\ Chem.\ Phys.\ {\bf 79}, 4412 (1983). 
\bibitem{farelly86}
	 D.\ Farelly: 
	 J. Chem. Phys. {\bf 85}, 2119 (1986).
\bibitem{bellissard+90}
	 J.\ Bellissard and M.\ Vittot: 
	 Ann. Inst. H. Poincar\'e {\bf 52}, 175 (1990).
\bibitem{eckhardt86}
	 B.\ Eckhardt:
	 J. Phys. A {\bf 19}, 2961 (1986).
\bibitem{bambusi95}
	 D.\ Bambusi: Nonlinearity {\bf 8}, 93 (1995).
\bibitem{private}
	 Private communication by J.\ Bellissard.
\bibitem{graffi+87}
	 S.\ Graffi and T.\ Paul: Comm.\ Math.\ Phys.\ {\bf 108}, 25 (1987).
\bibitem{jauslin93}
	 H.\ R.\ Jauslin: 
	 {\em Stability and Chaos in Classical and Quantum Hamiltonian 
	      Systems,\/}
	 in: {\em Computational Physics} 
	 P.\ L.\ Garrido and J.\ Marro (eds.), 
	 World Scientific 1993, p. 107.
%
\end{references}
\end{document}